\documentclass[twocolumn,aps,floats,showpacs,preprintnumbers]{revtex4}
\usepackage{epsfig}
\usepackage{amsmath}
\usepackage{amsfonts}
\usepackage{amssymb}

\def\nslash{\rlap{\hspace{0.02cm}/}{n}}
\def\vslash{\rlap{\hspace{0.02cm}/}{v}}

\def\Dslash{\rlap{\hspace{0.07cm}/}{D}}
\def\UVC{\Lambda_{\rm UV}}
\newcommand{\BM}{$B$-meson}
\newcommand{\UV}{$\Lambda_{\rm UV}$}

\begin{document}

\preprint{CLNS~05/1925}

\title{Model-Independent Properties of the B-Meson Distribution Amplitude}

\author{Seung J.~Lee$^a$ and Matthias Neubert$^{a,b}$}

\affiliation{$^a$Institute for High-Energy Phenomenology, Cornell University, 
Ithaca, NY 14853, USA\\
$^b$Institut f\"ur Theoretische Physik, Universit\"at Heidelberg, 
D--69120 Heidelberg, Germany}

\begin{abstract}
The operator product expansion is used to obtain model-independent predictions 
for the first two moments of the renormalized \BM\ light-cone distribution 
amplitude $\phi_+^B(\omega,\mu)$, defined with a cutoff $\omega\le\UVC$. The 
leading hadronic power corrections are given in terms of the parameter 
$\bar\Lambda=m_B-m_b$. From the cutoff dependence of the zeroth moment an 
analytical expression for the asymptotic behavior of the distribution 
amplitude is derived, which exhibits a negative radiation tail for 
$\omega\gg\mu$. By solving the evolution equation for the distribution 
amplitude, an integral representation for $\phi_+^B(\omega,\mu)$ is obtained 
in terms an initial function $\phi_+^B(\omega,\mu_0)$ defined at a lower 
renormalization scale. A realistic model of the \BM\ light-cone distribution 
amplitude is proposed, which satisfies the moment relations and has the 
correct asymptotic behavior. This model provides an estimate for the first 
inverse moment and the associated parameter $\lambda_B$.
\end{abstract}

\pacs{12.38.Cy,12.39.Hg,12.39.St,13.25.Hw} 
\maketitle

\section{Introduction}

Exclusive decays of $B$ mesons such as $B\to\pi l\nu$ and $B\to\pi\pi,\pi K$ 
are important tools to search for physics beyond the Standard Model as well as 
to measure fundamental parameters in the flavor sector. In processes where 
large momentum is transferred to the soft spectator quark via hard gluon 
exchange, the \BM\ light-cone distribution amplitude (LCDA) enters in the 
parameterization of hadronic matrix elements of bilocal current operators 
\cite{Grozin:1996pq}. The past few years have seen a lot of progress in the 
theoretical framework for the analysis of exclusive \BM\ decays, mainly based 
on QCD factorization theorems 
\cite{Beneke:1999br,Beneke:2000ry,Beneke:2001ev,Beneke:2003zv} and 
perturbative QCD methods 
\cite{Keum:2000ph,Keum:2000wi,Keum:2002cr,Keum:2003js}. However, in many cases 
the extraction of important physics from experimental data is still limited by 
theoretical uncertainties, often due to our ignorance of the functional form 
of the \BM\ LCDA and other hadronic matrix elements. For example, using the 
soft-collinear effective theory  
\cite{Bauer:2000yr,Bauer:2001yt,Chay:2002vy,Beneke:2002ph,Hill:2002vw}, the 
large-recoil heavy-to-light form factors relevant to weak $B$ decays have been 
studied at leading order in a $1/E$ expansion 
\cite{Bauer:2002aj,Beneke:2003pa,Lange:2003pk}. The analysis of spin-symmetry 
violating contributions to these form factors, in particular, relies on 
knowledge about the \BM\ LCDA 
\cite{Beneke:2000wa,Hill:2004if,Beneke:2004rc,Becher:2004kk}.

In spite of the importance of the \BM\ LCDA, so far most studies of its 
properties have been limited to model-dependent analyses based on QCD sum 
rules \cite{Grozin:1996pq,Ball:2003fq,Braun:2004}. In the present work, we 
employ the operator product expansion (OPE) to explore some model-independent 
properties of the LCDA. We calculate the first two moments of the distribution 
amplitude, derive its asymptotic behavior, and study its properties under 
renormalization-group evolution, thereby obtaining strong constraints on model 
building. Using the results of this analysis, we propose a realistic model of 
the \BM\ LCDA and use it to estimate the important hadronic parameter 
$\lambda_B$ \cite{Beneke:1999br}, which enters in many analyses based on QCD 
factorization.

\section{Moment analysis}
\label{sec:moments}

The leading-twist, two-particle LCDA $\phi_+^B$ of the \BM\ is defined in 
terms of the \BM\ matrix element of a renormalized bilocal heavy-quark 
effective theory (HQET) operator relative to the matrix element of the 
corresponding local operator. The bilocal operator is made up of a soft 
spectator quark $q_s$ and a heavy quark $h$ at light-like separation $z$, 
connected by a straight soft Wilson line $S_n(z,0)$. Specifically, one defines
\cite{Grozin:1996pq}
\begin{equation}\label{bilocal}
   \widetilde\phi_+^B(\tau,\mu)
   = \frac{\langle\,0\,|\,\bar q_s(z)\,S_n(z,0)\,\nslash\,\Gamma\,h(0)\,
           |\bar B(v)\rangle}%
          {\langle\,0\,|\,\bar q_s(0)\,\nslash\,\Gamma\,h(0)\,
           |\bar B(v)\rangle} \,,
\end{equation}
where $\tau=v\cdot z-i\epsilon$. Our notation is such that $z$ is proportional 
to a light-like vector $n$, $v$ is the \BM\ velocity, and for convenience we 
choose $n\cdot v=1$. The object $\Gamma$ represents an arbitrary Dirac matrix 
chosen such that the operators have nonzero overlap with the $B$ meson. The 
momentum-space LCDA is given by the Fourier transform
\begin{equation}
   \phi_+^B(\omega,\mu) = \frac{1}{2\pi} \int d\tau\,e^{i\omega\tau}\,
   \widetilde\phi_+^B(\tau,\mu) \,.
\end{equation}
The analytic properties of the function $\widetilde\phi_+^B(\tau,\mu)$ in the 
complex $\tau$ plane imply that $\phi_+^B(\omega,\mu)=0$ if $\omega<0$.

We start by defining regularized moments of the \BM\ LCDA as (for integer 
$N\ge 0$)
\begin{equation}\label{moment}
   M_N(\Lambda_{\rm UV},\mu) = \int_0^{\Lambda_{\rm UV}}\!
   d\omega\,\omega^N \phi_+^B(\omega,\mu) \,.
\end{equation}
A hard cutoff \UV\ is imposed on the integral so as to avoid singularities 
from the region of large $\omega$ values, which are not regularized by 
renormalizing the bilocal operator in (\ref{bilocal}) \cite{Grozin:1996pq}. 
The reason is that the position-space LCDA $\widetilde\phi_B^+(\tau,\mu)$ and 
its derivatives are singular at $\tau=0$. Only cut moments of the renormalized 
LCDA are UV finite. For a sufficiently large value of \UV\, the moments 
$M_N(\Lambda_{\rm UV},\mu)$ can be expanded in a series of \BM\ matrix 
elements of local HQET operators. The basic idea is the same as that used in 
previous work on cut moments of the $B$-meson shape function entering the 
analysis of inclusive decays \cite{Gil:2004,Bauer:2003pi}. From the structure 
of the bilocal HQET operator in (\ref{bilocal}) and the Feynman rules of HQET 
it follows that the resulting local operators have Dirac structure 
\begin{equation}
   \bar q_s\,(\gamma\gamma\dots\gamma)\,\nslash\,\Gamma\,h \,,
\end{equation}
where the number of Dirac matrices inside the parenthesis is even if light 
quarks are treated as massless. By using the equations of motion 
$i\Dslash\,q_s=0$ and $iv\cdot D\,h=0$, it is straightforward to find the 
corresponding operators of a given dimension $D$. For $D=3$, the only 
possibility is the operator
\begin{equation}
   Q_0 = \bar q_s\,\nslash\,\Gamma\,h \,,
\end{equation}
which appears in the denominator in (\ref{bilocal}). For $D=4$, there are 
naively four subleading operators with one derivative, namely
\begin{eqnarray}
   Q_{1a} &=& \bar q_s\,iv\cdot\overleftarrow{D}\,\nslash\,\Gamma\,h \,,
    \hspace{0.73cm}
   Q_{1c} = \bar q_s\,i\,n\cdot D\,\nslash\,\Gamma\,h \,, \nonumber\\
   Q_{1b} &=& \bar q_s\,in\cdot\overleftarrow{D}\,\nslash\,\Gamma\,h \,, 
    \qquad
   Q_{1d} = \bar q_s\,i\vslash\,\Dslash\,\nslash\,\Gamma\,h \,. \quad
\end{eqnarray}
However, the Wilson coefficients of the operators $Q_{1c}$ and $Q_{1d}$ are 
zero, because the residual momentum $k$ of the external heavy-quark field only 
appears as $v\cdot k$ in HQET diagrams. Hence, these operators can be ignored.
For $D\geq 5$, the situation becomes more complicated, since operators 
containing the gluon field $G^{\mu\nu}$ need to be included. For our current 
analysis, we restrict the discussion to operators of dimension less than 5. 

\begin{figure}
\epsfxsize=8.0cm 
\centerline{\epsffile{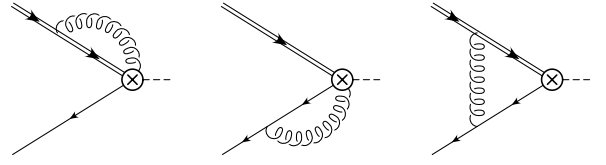}}
\caption{\label{fig:graphs} 
One-loop diagrams contributing to the partonic matrix elements of bilocal and 
local operators in HQET. A crossed circle denotes an operator insertion. 
Double lines represent effective heavy-quark fields.}
\end{figure}

The resulting expansion of the moments to subleading power in $1/\UVC$ takes 
the form
\begin{eqnarray}\label{momentsexpansion}
   M_N(\UVC,\mu)
   &=& \UVC^N\,\Bigg\{ K_0^{(N)}(\UVC,\mu) \nonumber\\
   &&\hspace{-2.2cm}\mbox{}+ \sum_{i=a,b} \frac{K_{1i}^{(N)}(\UVC,\mu)}{\UVC}\,
    \frac{\langle\,0\,|\,Q_{1i}\,|\bar B(v)\rangle}%
         {\langle\,0\,|\,Q_0\,|\bar B(v)\rangle} + \dots \Bigg\} \,, \quad
\end{eqnarray}
where the ellipses denote terms of order $(\Lambda_{\rm QCD}/\UVC)^2$ and 
higher. The short-distance coefficients $K_n^{(N)}(\UVC,\mu)$ can be 
calculated using on-shell external quark states and employing partonic 
expressions for the LCDA and for the matrix elements of the local operators 
$Q_n$ to evaluate both sides of the matching relation 
(\ref{momentsexpansion}). The relevant one-loop diagrams are shown in 
Figure~\ref{fig:graphs}. Wave-function renormalization contributions cancel in 
the matching and thus can be omitted. We assign incoming residual momentum $k$ 
to the heavy quark and incoming momentum $p$ to the light quark, subject to 
the on-shell conditions $v\cdot k=0$ and $p^2=0$. The Feynman amplitude is 
expanded to linear order in $p$ before loop integrations are performed. This 
ensures that loop corrections to the matrix elements of the local operators 
vanish in dimensional regularization, because all integrals are scaleless. We 
thus obtain
\begin{eqnarray}\label{Qiparton}
   \langle\,0\,|\,Q_{1a}\,|\bar B(v)\rangle_{\rm parton}
   &=& v\cdot p\,\langle\,0\,|\,Q_0\,|\bar B(v)\rangle_{\rm parton} \,,
    \nonumber\\
   \langle\,0\,|\,Q_{1b}\,|\bar B(v)\rangle_{\rm parton}
   &=& n\cdot p\,\langle\,0\,|\,Q_0\,|\bar B(v)\rangle_{\rm parton} \,. \quad
\end{eqnarray}

\noindent
The result for the one-loop matrix element of the bilocal HQET operator is 
nontrivial. According to (\ref{bilocal}), it provides us with a partonic 
expression for the LCDA. After $\overline{\rm MS}$ subtractions, we obtain at 
one-loop order
\begin{eqnarray}\label{phiparton}
   \phi^B_+(\omega,\mu)_{\rm parton}
   &=& \delta(\omega) \left( 1 - \frac{C_F\alpha_s}{4\pi}\,
    \frac{\pi^2}{12} \right) \\
   &&\hspace{-2.5cm}\mbox{}+ \frac{C_F\alpha_s}{4\pi} \left[
    - 4 \left( \frac{\ln{\frac{\omega}{\mu}}}{\omega} \right)_*^{\![\mu]} 
    + 2 \left( \frac{1}{\omega} \right)_*^{\![\mu]} \right] \nonumber\\
   &&\hspace{-2.5cm}\mbox{}+ \delta'(\omega) \left\{ 
    - n\cdot p \left[ 1 - \frac{C_F\alpha_s}{4\pi}
    \left( 1 + \frac{\pi^2}{12} \right) \right]
    + v\cdot p\,\frac{C_F\alpha_s}{4\pi} \right\} \nonumber\\
   &&\hspace{-2.5cm}\mbox{}+ \frac{C_F\alpha_s}{4\pi}
    \left[ - 4 n\cdot p 
    \left( \frac{\ln{\frac{\omega}{\mu}}}{\omega^2} \right)_*^{\![\mu]} 
    \!\! + (5 n\cdot p + 4 v\cdot p)
    \left( \frac{1}{\omega^2} \right)_*^{\![\mu]} \right] \! , \nonumber
\end{eqnarray}
where $\alpha_s\equiv\alpha_s(\mu)$ throughout, unless indicated otherwise. We 
have retained terms of linear order in $p$, which will be sufficient to 
extract the matching coefficients $K_0^{(N)}$ and $K_{1i}^{(N)}$ in 
(\ref{momentsexpansion}). The star distributions are generalized plus 
distributions defined as
\begin{eqnarray}
   \int_0^\Lambda\!d\omega\,F_\omega\!
    \left( \frac{1}{\omega} \right)_*^{\![\mu]}
   &=& \int_0^\Lambda\!d\omega\,\frac{F_\omega-F_0}{\omega}
    + F_0\,\ln{\frac{\Lambda}{\mu}} \,, \nonumber\\
   \int_0^\Lambda\!d\omega\,F_\omega\!
    \left( \frac{\ln{\frac{\omega}{\mu}}}{\omega} \right)_*^{\![\mu]}
   &=& \int_0^\Lambda\!d\omega\,\frac{F_\omega-F_0}{\omega}\,
    \ln{\frac{\omega}{\mu}}
    + \frac{F_0}{2}\,\ln^2{\frac{\Lambda}{\mu}} \,, \nonumber\\
   \int_0^\Lambda\!d\omega\,F_\omega\!
    \left( \frac{1}{\omega^2} \right)_*^{\![\mu]}
   &=& \int_0^\Lambda\!d\omega\,\frac{F_\omega-F_0-\omega\,F'_0}{\omega^2}
    \nonumber\\
   &&\mbox{}- \frac{F_0}{\Lambda} + F'_0\,\ln{\frac{\Lambda}{\mu}} \,, \\
   \int_0^\Lambda\!d\omega\,F_\omega\!
    \left( \frac{\ln{\frac{\omega}{\mu}}}{\omega^2} \right)_*^{\![\mu]}
   &=& \int_0^\Lambda\!d\omega\,\frac{F_\omega-F_0-\omega\,F'_0}{\omega^2}
    \,\ln{\frac{\omega}{\mu}} \nonumber\\
   &&\mbox{}- \frac{F_0}{\Lambda} \left( \ln\frac{\Lambda}{\mu} + 1 \right)
    + \frac{F'_0}{2}\,\ln^2{\frac{\Lambda}{\mu}} \,, \nonumber
\end{eqnarray}
where $F(\omega)$ is a smooth test function, and we use the short-hand 
notation $F_\omega\equiv F(\omega)$ and $F'_\omega\equiv F'(\omega)$.

Given the results (\ref{Qiparton}) and (\ref{phiparton}), it is 
straightforward to derive expressions for both sides of the matching relation 
(\ref{momentsexpansion}) in the parton model, and to extract the desired 
expressions for the Wilson coefficients. At one-loop order, we find
\begin{eqnarray}
   K_0^{(0)} &=& 1 + \frac{C_F\alpha_s}{4\pi} \left( 
    - 2\ln^2\frac{\UVC}{\mu} + 2\ln\frac{\UVC}{\mu} - \frac{\pi^2}{12}
    \right) , \nonumber\\
   K_{1a}^{(0)} &=& \frac{C_F\alpha_s}{4\pi}\,(-4) \,, \nonumber\\
   K_{1b}^{(0)} &=& \frac{C_F\alpha_s}{4\pi} \left( 4\ln\frac{\UVC}{\mu} - 1
    \right)
\end{eqnarray}
for the zeroth moment, and
\begin{eqnarray}
   K_0^{(1)} &=& \frac{C_F\alpha_s}{4\pi} \left( - 4\ln\frac{\UVC}{\mu}
    + 6 \right) , \nonumber\\
   K_{1a}^{(1)} &=& \frac{C_F\alpha_s}{4\pi} \left( 4\ln\frac{\UVC}{\mu} - 1
    \right) , \\
   K_{1b}^{(1)} &=& 1 + \frac{C_F\alpha_s}{4\pi} \left( 
    - 2\ln^2\frac{\UVC}{\mu} + 5\ln\frac{\UVC}{\mu} - 1 - \frac{\pi^2}{12}
    \right) \nonumber
\end{eqnarray}
for the first moment. 

We have repeated the entire calculation outlined above in a different  
regularization scheme, where the dependence of the Feynman amplitudes on the 
component $n\cdot p$ of the light-quark momentum is kept exactly, whereas we 
linearize in the remaining components of $p$. In this scheme the loop 
corrections to the matrix elements of the local operators in (\ref{Qiparton}) 
no longer vanish, and the result for the LCDA is far more complicated than 
that displayed in (\ref{phiparton}). Nevertheless, we obtain the same 
expressions for the Wilson coefficients $K_n^{(0)}$ and $K_n^{(1)}$ as given 
above. This is a highly nontrivial check, which gives us confidence in the 
correctness of our results.

The Wilson coefficients describe the short-distance physics associated with 
the large cutoff scale \UV, and hence it was legitimate to obtain them using a 
partonic calculation. Long-distance effects, on the other hand, reside in the 
hadronic matrix elements of the local operators $Q_n$, which cannot be 
calculated reliably using perturbation theory. However, these matrix elements 
are constrained by heavy-quark symmetry and can be parameterized in terms of 
universal form factors \cite{Neubert:1993mb}. The results are particularly 
simple in the case of the operators $Q_{1i}$. Using relations derived in 
\cite{Neubert:1992c}, we find that
\begin{equation}
   \frac{\langle\,0\,|\,Q_{1a}\,|\bar B(v)\rangle}%
        {\langle\,0\,|\,Q_0\,|\bar B(v)\rangle} = \bar\Lambda \,, \qquad
   \frac{\langle\,0\,|\,Q_{1b}\,|\bar B(v)\rangle}%
        {\langle\,0\,|\,Q_0\,|\bar B(v)\rangle} = \frac{4\bar\Lambda}{3} \,,
\end{equation}
where the quantity $\bar\Lambda=m_B-m_b$ is the only hadronic parameter needed 
at this order. The first-order power corrections to the moments $M_N$ can now 
be expressed in terms of $\bar\Lambda$. At one-loop order, and to subleading 
order in the power expansion in $1/\UVC$, the results are
\begin{eqnarray}\label{moments2}
   M_0 &=& 1 + \frac{C_F\alpha_s}{4\pi} \left(
    - 2\ln^2\frac{\UVC}{\mu} + 2\ln\frac{\UVC}{\mu} - \frac{\pi^2}{12} \right)
    \nonumber\\
   &&\mbox{}+ \frac{16\bar\Lambda}{3\UVC}\,
    \frac{C_F\alpha_s}{4\pi} \left( \ln\frac{\UVC}{\mu} - 1 \right) ,
    \nonumber\\
   M_1 &=& \UVC\,\frac{C_F\alpha_s}{4\pi} \left(
    - 4\ln\frac{\UVC}{\mu} + 6 \right) \\
   &&\hspace{-1.0cm}\mbox{}+ \frac{4\bar\Lambda}{3} \left[
    1 + \frac{C_F\alpha_s}{4\pi} \left(
    - 2\ln^2\frac{\UVC}{\mu} + 8\ln\frac{\UVC}{\mu} - \frac74
    - \frac{\pi^2}{12} \right) \right] \! . \nonumber
\end{eqnarray}
These are our final expressions for the first two moments of the renormalized 
\BM\ LCDA. As long as $\UVC\gg\Lambda_{\rm QCD}$, they are model-independent 
predictions of QCD, valid up to higher-order terms in $\alpha_s$ and $1/\UVC$. 
The fixed-order perturbative expressions derived here are applicable if the 
two scales $\UVC$ and $\mu$ are of the same order, so that the logarithms in 
the matching coefficients are not parametrically large.

Taking the derivative of the zeroth moment $M_0$ in (\ref{moments2}) with 
respect to the cutoff, we can obtain a model-independent description of the 
asymptotic behavior of the \BM\ LCDA \cite{Gil:2004}, i.e.\
\begin{equation}
   \phi_+^B(\omega,\mu) = \frac{dM_0(\UVC,\mu)}{d\UVC}\bigg|_{\UVC=\omega} \,.
\end{equation}
At one-loop order, the result reads
\begin{equation}\label{tail}
   \phi_+^B(\omega,\mu)
   = \frac{C_F\alpha_s}{\pi\omega} \! \left[ 
   \left( \frac12 - \ln\frac{\omega}{\mu} \right) \!
   + \frac{4\bar\Lambda}{3\omega} \!
   \left ( 2 - \ln\frac{\omega}{\mu} \right) \! + \dots \right] \! .
\end{equation}
This relation holds for $\omega\gg\Lambda_{\rm QCD}$, up to power corrections
of order $\Lambda_{\rm QCD}^2/\omega^3$. We observe that the radiation tail of 
the \BM\ LCDA becomes negative at $\omega\approx\sqrt{e}\mu$ for a 
sufficiently large value of $\mu$. This model-independent prediction for the 
asymptotic behavior of $\phi_+^B(\omega,\mu)$ agrees qualitatively with the 
findings of the QCD sum-rule analysis in \cite{Braun:2004}.

\section{Elimination of the pole mass}

Our calculations so far have been performed in the on-shell (pole) scheme, 
where $\bar\Lambda=m_B-m_b^{\rm pole}$ is defined in terms of the $b$-quark 
pole mass. However, it is well known that the pole mass suffers from infrared 
renormalon ambiguities \cite{Bigi:1994,Beneke:1994}. Hence, it is desirable to
eliminate the pole-scheme parameter $\bar\Lambda$ in favor of a new, 
short-distance parameter $\bar\Lambda_{\rm RS}$ defined in some 
renormalization scheme. For our purposes it is most convenient to employ a 
so-called ``low-scale subtracted'' heavy-quark mass defined with the help of a 
hard subtraction scale $\mu_f$. Examples are the ``kinetic mass'' 
\cite{Bigi:1996si}, the ``potential-subtracted mass'' \cite{Beneke:1998rk}, 
the ``1S mass'' \cite{Hoang:1998hm}, and the ``shape-function mass''
\cite{Gil:2004,Neubert:2004sp}. Using the last definition as an example, we 
would use the relation
\begin{equation}\label{LSF}
   \bar\Lambda = \bar\Lambda_{\rm SF}(\mu_f,\mu)
   + \mu_f\,\frac{C_F\alpha_s}{4\pi}
   \left( 8\ln\frac{\mu_f}{\mu} - 4 \right) + \dots
\end{equation}
to eliminate the pole-scheme parameter $\bar\Lambda$ in the moment relations 
(\ref{moments2}), identifying the subtraction scale $\mu_f$ with the cutoff 
$\UVC$. As always, $\alpha_s\equiv\alpha_s(\mu)$.

Alternatively, the moment relations themselves can be used to define a new 
subtraction scheme. Guided by the tree-level relations $M_1=4\bar\Lambda/3$ 
and $M_0=1$, we are led to define a running parameter (the subscript ``DA'' 
stands for ``distribution amplitude'')
\begin{equation}\label{LDAdef}
   \bar\Lambda_{\rm DA}(\mu_f,\mu)
   \equiv \frac{3M_1(\mu_f,\mu)}{4M_0(\mu_f,\mu)}
\end{equation}
to all orders in perturbation theory. From (\ref{moments2}), it follows that
\begin{eqnarray}\label{LDA}
   \bar\Lambda
   &=& \bar\Lambda_{\rm DA}(\mu_f,\mu)
    \left[ 1 - \frac{C_F\alpha_s}{4\pi}
    \left( 6\ln\frac{\mu_f}{\mu} - \frac74 \right) \right] \nonumber\\
   &&\mbox{}+ \mu_f\,\frac{C_F\alpha_s}{4\pi}
    \left( 3\ln\frac{\mu_f}{\mu} - \frac92 \right) + \dots \,.
\end{eqnarray}
By taking the ratio of $M_1$ and $M_0$ in (\ref{LDAdef}) the 
double-logarithmic radiative corrections are eliminated. Like the other 
short-distance mass definitions mentioned above, the parameter 
$\bar\Lambda_{\rm DA}$ can be regarded as a ``physical'' quantity in the sense 
that it is free of renormalon ambiguities. Perturbative relations can be used 
to transform from our new scheme to any other mass-definition scheme. For 
example, from (\ref{LSF}) and (\ref{LDA}) it follows that at one-loop order 
the parameter $\bar\Lambda_{\rm DA}$ is related to the parameter 
$\bar\Lambda_{\rm SF}$ in the shape-function scheme through
\begin{eqnarray}\label{Lbarrel}
   \bar\Lambda_{\rm DA}(\mu_f,\mu)
   &=& \bar\Lambda_{\rm SF}(\mu_*,\mu_*)
    \left[ 1 + \frac{C_F\alpha_s}{4\pi}
    \left( 6\ln\frac{\mu_f}{\mu} - \frac74 \right) \right] \nonumber\\
   &&\mbox{}- \mu_f\,\frac{C_F\alpha_s}{4\pi}
    \left( 3\ln\frac{\mu_f}{\mu} - \frac92 + \frac{4\mu_*}{\mu_f} \right) .
\end{eqnarray}

A rather precise value for $\bar\Lambda_{\rm SF}$ has been extracted from 
moment analyses of various spectra in the inclusive decays $B\to X_s\gamma$ 
and $B\to X_u l\,\nu$, yielding 
$\bar\Lambda_{\rm SF}(\mu_*,\mu_*)=(0.65\pm 0.06)$\,GeV at $\mu_*=1.5$\,GeV 
(and at leading order in $1/m_b$) \cite{Neubert:2004sp,Neubert:2005nt}. This 
value will be used as an input when we compute the running parameter 
$\bar\Lambda_{\rm DA}(\mu_f,\mu)$ from the above relation.

\section{Renormalization-group evolution}

In Section~\ref{sec:moments} we have derived model-independent predictions for 
moments of the \BM\ LCDA and for its asymptotic behavior for large $\omega$. 
The renormalization group can be used to obtain a model-independent 
description of how $\phi_+^B(\omega,\mu)$ changes under variation of the scale 
$\mu$. The integro-differential evolution equation obeyed by the LCDA was 
derived in \cite{Bjorn:2003}, where an analytic solution was presented in the 
form of a double integral. One finds that the distribution amplitude at a 
scale $\mu$ can be expressed in terms of that at a lower scale $\mu_0<\mu$ by
\begin{equation}\label{phif}
   \phi_+^B(\omega,\mu)
   = \frac{1}{2\pi} \int_{-\infty}^\infty\!dt\,\varphi_0(t)\,
   f(\omega,\mu,\mu_0,it) \,,
\end{equation}
where
\begin{equation}
   \varphi_0(t) = \int^{\infty}_{0}\!\frac{d\omega'}{\omega'}\,
   \phi_+^B(\omega',\mu_0) \left( \frac{\omega'}{\mu_0} \right)^{-it}
\end{equation}
denotes the Fourier transform with respect to $\ln\omega$ of the function 
$\phi_+^B(\omega,\mu_0)$ at the initial scale $\mu_0$. At leading order in 
perturbation theory, the kernel $f$ takes the form
\begin{eqnarray}\label{functionf}
   f(\omega,\mu,\mu_0,it)
   &=& e^{V(\mu,\mu_0)}  \left( \frac{\omega}{\mu_0} \right)^{it+g}
    e^{-2\gamma_E g} \nonumber\\
   &\times& \frac{\Gamma(1-it-g)\,\Gamma(1+it)}{\Gamma(1+it+g)\,\Gamma(1-it)}
    \,,
\end{eqnarray}
where 
\begin{equation}
   V(\mu,\mu_0)
   = - \int\limits_{\alpha_s(\mu_0)}^{\alpha_s(\mu)}\!\!
   \frac{d\alpha}{\beta(\alpha)} \Bigg[ \Gamma_{\rm cusp}(\alpha)
   \!\!\int\limits_{\alpha_s(\mu_0)}^\alpha\!\!
   \frac{d\alpha'}{\beta(\alpha')} + \gamma(\alpha) \Bigg] \,,
\end{equation}
and
\begin{equation}
   g\equiv g(\mu,\mu_0)
   = \int\limits_{\alpha_s(\mu_0)}^{\alpha_s(\mu)}\!\!\!d\alpha\,
    \frac{\Gamma_{\rm cusp}(\alpha)}{\beta(\alpha)}
   \approx \frac{2C_F}{\beta_0} \ln\frac{\alpha_s(\mu_0)}{\alpha_s(\mu)} \,.
\end{equation}
In these expressions $\beta=d\alpha_s/d\ln\mu$ is the $\beta$-function, and 
$\Gamma_{\rm cusp}=C_F\alpha_s/\pi+\dots$, $\gamma=-C_F\alpha_s/2\pi+\dots$ 
are anomalous dimensions. The perturbative expansion of $V(\mu,\mu_0)$ at 
next-to-leading order can be found in \cite{Bosch:2003fc}.

Here we take a step further and simplify the solution obtained in 
\cite{Bjorn:2003} by performing the integration over $t$ in (\ref{phif}) 
analytically. Substituting the expression for $f$ from (\ref{functionf}), we 
observe that the integrand has poles situated on the imaginary axis in the 
complex $t$ plane. The poles on the negative imaginary axis are located at 
$t=-i(n-g)$ with $n\ge 1$ an integer (we assume $0<g<1$, which is satisfied 
for all reasonable values of scales), while those on the positive imaginary 
axis are located at $t=in$ with $n\ge 1$ an integer. Using the theorem of 
residues, we obtain
\begin{eqnarray}\label{nice}
   \phi_+^B(\omega,\mu)
   &=& e^{V(\mu,\mu_0)}\,e^{-2\gamma_E g}\,\frac{\Gamma(2-g)}{\Gamma(g)}
    \int_0^\infty\!\frac{d\omega'}{\omega'}\,\phi_+^B(\omega',\mu_0)
    \nonumber\\
   &\times& \left( \frac{\omega_>}{\mu_0} \right)^g \frac{\omega_<}{\omega_>}  
    \,{}_2F_1\Big(1-g,2-g; 2; \frac{\omega_<}{\omega_>}\Big) \,,
\end{eqnarray}
where $\omega_<=\min(\omega,\omega')$ and $\omega_>=\max(\omega,\omega')$. The 
hypergeometric function ${}_2F_1(a,b; c; z)$ has the series expansion
\begin{equation}
   {}_2F_1(a,b; c; z) = \sum_{n=0}^\infty
   \frac{(a)_n\,(b)_n}{(c)_n}\,\frac{z^n}{n!}
\end{equation}
with $(a)_n=\Gamma(a+n)/\Gamma(a)$. In the limit $\mu\to\mu_0$ we have 
$V(\mu,\mu_0)\to 0$, $g\to 0$, and ${}_2F_1(1-g,2-g; 2; x)\to (1-x)^{2g-1}$. 
Then the right-hand side in (\ref{nice}) reduces to the left-hand one.

Equation~(\ref{nice}) provides the most compact expression possible for 
calculating the evolution of the LCDA under changes of the renormalization 
scale. It is tempting to conjecture that this is the exact solution to the 
evolution equation for the LCDA, valid to all orders in perturbation theory. 
An analogous statement is indeed true for the $B$-meson shape function 
\cite{Neubert:2004dd}. In the present case, to prove this assertion one would 
need to show that the exact evolution equation for the LCDA is given by 
\begin{eqnarray}
   && \left[ \frac{d}{d\ln\mu}
    + \Gamma_{\rm cusp}(\alpha_s)\,\ln\frac{\mu}{\omega}
    + \gamma(\alpha_s) \right] \phi_+^B(\omega,\mu) \nonumber\\
   &&= \Gamma_{\rm cusp}(\alpha_s)
    \int_0^\infty\!d\omega'\,\frac{\omega}{\omega_>}\,
    \frac{\phi_+^B(\omega',\mu)-\phi_+^B(\omega,\mu)}{|\omega'-\omega|} \,,
    \qquad 
\end{eqnarray}
where $\Gamma_{\rm cusp}$ is the universal cusp anomalous dimension of Wilson 
loops with light-like segments \cite{Korchemsky:wg,Korchemskaya:1992je}, and 
$\gamma$ is some other anomalous dimension. In \cite{Bjorn:2003}, the above 
relation was confirmed at one-loop order.

\section{Phenomenological model}

The model-independent properties of the \BM\ distribution amplitude derived in 
this work provide useful constraints on model building. In this section we 
suggest a realistic form for $\phi_+^B(\omega,\mu)$, which satisfies these 
constraints. For phenomenological purposes such a model is needed at a 
renormalization scale of order $\mu\sim\sqrt{m_b\Lambda_{\rm QCD}}$, as this 
is the characteristic ``hard-collinear'' scale for hard spectator scattering 
in exclusive $B$ decays \cite{Hill:2002vw,Bosch:2003fc}. Our model consists of 
the two-component ansatz
\begin{eqnarray}\label{BLCDAmodel}
   \phi_+^B(\omega,\mu)
   &=& N\,\frac{\omega}{\omega_0^2}\,
    e^{-\omega/\omega_0} + \theta(\omega-\omega_t)\,
    \frac{C_F\alpha_s}{\pi\omega} \nonumber\\
   &\times& \left[ \left( \frac12 - \ln\frac{\omega}{\mu} \right)
    + \frac{4\bar\Lambda_{\rm DA}}{3\omega}
    \left( 2 - \ln\frac{\omega}{\mu} \right) \right] , \quad
\end{eqnarray}
where $\bar\Lambda_{\rm DA}\equiv\bar\Lambda_{\rm DA}(\mu,\mu)$ is defined in 
our new scheme (\ref{LDA}), and we set $\mu_f=\mu$ for simplicity. The first 
term on the right-hand side is based on the exponential form proposed in 
\cite{Grozin:1996pq}, while the second piece is a radiation tail added so as 
to ensure the correct asymptotic behavior as shown in (\ref{tail}). The tail 
is ``glued'' onto the exponential at a position $\omega_t$ chosen such that 
the resulting function is continuous. This yields
\begin{equation}\label{wt}
   \omega_t = \sqrt{e}\mu \left( 1 + \frac{2\bar\Lambda_{\rm DA}}{\sqrt{e}\mu}
   - \frac{14\bar\Lambda_{\rm DA}^2}{3e\mu^2} + \dots \right) .
\end{equation}
The normalization constant $N$ and the parameter $\omega_0$ can be fixed by 
matching the expressions for the first two moments in (\ref{moments2}) with 
the corresponding results obtained by substituting the model function 
(\ref{BLCDAmodel}) into (\ref{moment}), neglecting exponentially small terms 
$\sim e^{-\Lambda_{\rm UV}/\omega_0}$. All remaining terms involving the 
cutoff \UV\ are reproduced by construction, so that the results for $N$ and 
$\omega_0$ are independent of the cutoff, as they must be. At first order in 
$\alpha_s$, we obtain
\begin{eqnarray}\label{Nconst}
   N &=& 1 + \frac{C_F\alpha_s}{4\pi} \Bigg[
    - 2\ln^2\frac{\omega_t}{\mu} + 2\ln\frac{\omega_t}{\mu} - \frac{\pi^2}{12}
    \nonumber\\
   &&\mbox{}+ \frac{16\bar\Lambda_{\rm DA}}{3\omega_t}
    \left( \ln\frac{\omega_t}{\mu} - 1 \right) \Bigg] \nonumber\\
   &=& 1 + \frac{C_F\alpha_s}{4\pi} \left( \frac12 - \frac{\pi^2}{12}
    - \frac{8\bar\Lambda_{\rm DA}}{3\sqrt{e}\mu} + \dots \right) ,
\end{eqnarray}
and
\begin{eqnarray}\label{w0}
   \omega_0 &=& \frac{2\bar\Lambda_{\rm DA}}{3} \left\{ 1
    +\! \frac{C_F\alpha_s}{4\pi} \left[ 6\ln\frac{\omega_t}{\mu}
    - \!\frac{16\bar\Lambda_{\rm DA}}{3\omega_t}\!
    \left( \ln\frac{\omega_t}{\mu} - 1 \right) \right] \! \right\} \nonumber\\
   &&\mbox{}- \frac{C_F\alpha_s}{4\pi} \left[ \omega_t
    \left( 2\ln\frac{\omega_t}{\mu} - 3 \right) + 3\mu \right] \\
   &=& \frac{2\bar\Lambda_{\rm DA}}{3} \left( 1
    + 3\,\frac{C_F\alpha_s}{4\pi} \right)
    + (2\sqrt{e}-3) \mu\,\frac{C_F\alpha_s}{4\pi} + \dots \,. \nonumber
\end{eqnarray}
The expanded expressions for $\omega_t$, $N$, and $\omega_0$ are given only 
for the purpose of illustration. The exact expressions will be used in our 
numerical analysis.

\begin{table}
\caption{\label{tab:pars}
Parameters of the model function (\ref{BLCDAmodel}) for different values of 
the renormalization scale}
\vspace{0.1cm}
\begin{center}
\begin{tabular}{|c|cccc|}
\hline\hline
$\mu$~[GeV] & $\bar\Lambda_{\rm DA}$~[GeV] & $\omega_t$~[GeV] & $N$
 & $\omega_0$~[GeV] \\
\hline
1.0 & 0.519 & 2.33 & 0.963 & 0.438 \\
1.5 & 0.635 & 3.35 & 0.974 & 0.509 \\
2.0 & 0.709 & 4.32 & 0.978 & 0.557 \\
2.5 & 0.770 & 5.26 & 0.981 & 0.596 \\
\hline\hline
\end{tabular}
\end{center}
\end{table}

\begin{figure}
\epsfxsize=8.0cm
\centerline{\epsffile{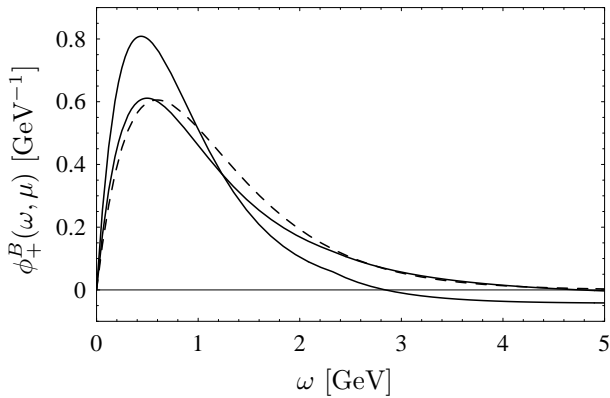}}
\caption{\label{fig:funs} 
Model ansatz for the \BM\ LCDA at $\mu=1$\,GeV (narrow solid curve) and 
2.5\,GeV (wide solid curve). The dashed curve shows the result at 2.5\,GeV 
obtained by evolving the distribution amplitude from 1 to 2.5\,GeV.}
\end{figure}

The model ansatz (\ref{BLCDAmodel}) has the attractive feature that it is to a 
good approximation invariant under renormalization-group evolution. 
Table~\ref{tab:pars} collects the parameters entering this function for 
different values of $\mu$, obtained using the central value 
$\bar\Lambda_{\rm SF}(\mu_*,\mu_*)=0.65$\,GeV in (\ref{Lbarrel}). For 
$\mu=1$\,GeV and 2.5\,GeV the corresponding functions are shown in 
Figure~\ref{fig:funs}. For comparison, we also show the result at 
$\mu=2.5$\,GeV obtained by applying the evolution formula (\ref{nice}) to the 
model function at $\mu=1$\,GeV. Both curves are very similar, indicating that 
the functional form (\ref{BLCDAmodel}) is approximately preserved under 
evolution.

\begin{figure}
\epsfxsize=8.0cm
\centerline{\epsffile{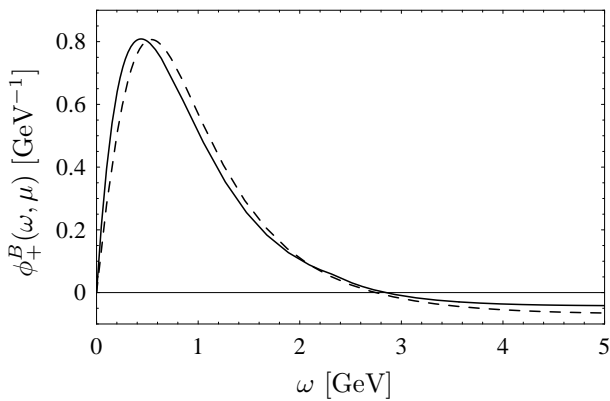}}
\caption{\label{fig:comparison} 
Two different models for the \BM\ LCDA at $\mu=1$\,GeV, constrained to have 
the same normalization and first moment. The solid curve corresponds to 
(\ref{BLCDAmodel}), the dashed one to (\ref{phiBIK}).}
\end{figure}

We have mentioned earlier that a QCD sum-rule analysis of the \BM\ LCDA at 
next-leading order in $\alpha_s$ performed by Braun et al.\ \cite{Braun:2004} 
has exhibited an asymptotic behavior similar to that of our perturbative QCD 
analysis. These authors have proposed the model form
\begin{equation}\label{phiBIK}
   \phi_+^B(\omega,\mu)
   = \frac{4\lambda_B^{-1}}{\pi}\,\frac{k}{k^2+1}
   \left[ \frac{1}{k^2+1} - \frac{2(\sigma_B-1)}{\pi^2}\,\ln k \right]
\end{equation}
at $\mu=1$\,GeV, where $k=\omega/1$\,GeV. The two parameters entering this 
functions are defined in terms of the integrals
\begin{eqnarray}\label{inverse}
   \lambda_B^{-1} &=& \int_0^\infty\!d\omega\,
    \frac{\phi_+^B(\omega,\mu)}{\omega} \,, \nonumber\\
   \sigma_B\,\lambda_B^{-1} &=& - \int_0^\infty\!d\omega\,
    \frac{\phi_+^B(\omega,\mu)}{\omega}\ln\frac{\omega}{\mu}\, \,.
\end{eqnarray} 
The parameter ranges obtained from the sum-rule analysis are 
$\lambda_B^{-1}=(2.15\pm 0.50)$\,GeV$^{-1}$ and $\sigma_B=1.4\pm 0.4$ at 
$\mu=1$\,GeV. On the other hand, if we require that the function 
(\ref{phiBIK}) obey the moment constraints (\ref{moments2}) at a large value 
of the cutoff, say $\UVC=3$\,GeV, then we find 
$\lambda_B^{-1}=(1.79\pm 0.06)$\,GeV$^{-1}$ and $\sigma_B=1.57\pm 0.27$. These 
values are consistent with the findings of \cite{Braun:2004}. It is 
interesting that, once the moment constraints are imposed, the two models in 
(\ref{BLCDAmodel}) and (\ref{phiBIK}) are nearly indistinguishable, in spite 
of the rather different functional forms (exponential vs.\ power-like 
fall-off). This fact is illustrated in Figure~\ref{fig:comparison}.

\begin{figure}
\epsfxsize=7.0cm
\centerline{\epsffile{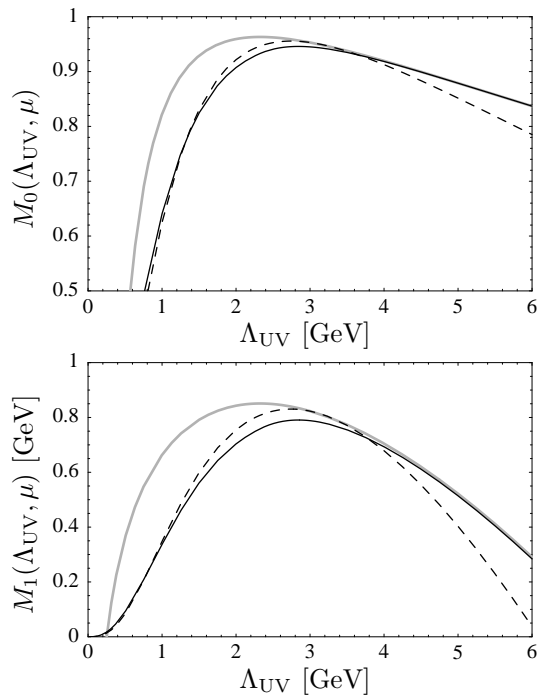}}
\caption{\label{fig:Mncomp} 
Comparison of model results (black) and OPE predictions (gray) for the first 
two moments of the LCDA, evaluated at $\mu=1$\,GeV and for different values of 
the cutoff. The solid black curves are obtained in our model 
(\ref{BLCDAmodel}), the dashed ones in the model (\ref{phiBIK}) of 
\cite{Braun:2004}.}
\end{figure}

We are now in a position to investigate how moments of the LCDA computed using 
the model functions (\ref{BLCDAmodel}) and (\ref{phiBIK}) compare with the 
model-independent predictions (\ref{moments2}) of the OPE, which are valid for
$\UVC\gg\Lambda_{\rm QCD}$. In Figure~\ref{fig:Mncomp}, we show in black the
model results for the moments $M_0$ and $M_1$ at $\mu=1$\,GeV as a function of 
the cutoff \UV. For comparison, the gray curves show the predictions of the 
OPE. We observe that our model curves quickly converge toward the OPE 
predictions for $\UVC>2.5$\,GeV. For large cutoff values the agreement is 
perfect, since by construction our function has the correct asymptotic 
behavior. The model of Braun et al.\ agrees qualitatively with the OPE for 
large \UV, but exact agreement can only be enforced at a single value of the 
cutoff (3\,GeV in our case). Note that for small values of \UV\ there are 
significant deviations between the OPE predictions and the model results. This 
is expected, given that the OPE is only valid for $\UVC\gg\Lambda_{\rm QCD}$. 
For $\UVC=2$\,GeV, for example, we expect unknown corrections of order 
$(\bar\Lambda/\UVC)^2\sim 0.1$ to $M_0$, and of order 
$\bar\Lambda^2/\UVC\sim 0.2$\,GeV to $M_1$. This is consistent with the 
deviations seen in the figure.

\section{Estimates for inverse moments}

The ``inverse moments'' defined in (\ref{inverse}) play an important role in 
the analysis of many exclusive $B$-meson decays. They control the strength of 
the leading-power spectator interactions in leptonic decays such as 
$B\to\gamma l\nu$, semileptonic decays such as $B\to\pi l\nu$, and hadronic 
decays such as $B\to\pi\pi$. The quantity $\sigma_B$ enters these analyses as 
soon as one goes beyond the tree approximation. Given that we have constructed 
highly constrained models for the distribution amplitude which satisfy the QCD 
predictions for moments and have the correct asymptotic behavior, it is 
interesting to ask what estimates we can obtain for the parameters $\lambda_B$ 
and $\sigma_B$.

\begin{table}
\caption{\label{tab:lamB}
Inverse moments $\lambda_B^{-1}$ and $\sigma_B$ calculated using the model 
function (\ref{BLCDAmodel})}
\vspace{0.1cm}
\begin{center}
\begin{tabular}{|c|cc|}
\hline\hline
$\mu$~[GeV] & $\lambda_B^{-1}$~[GeV$^{-1}$] & $\sigma_B$ \\
\hline
1.0 & $2.09\pm 0.24$ & $1.61\pm 0.09$ \\
1.5 & $1.86\pm 0.17$ & $1.79\pm 0.08$ \\
2.0 & $1.72\pm 0.14$ & $1.95\pm 0.07$ \\
2.5 & $1.62\pm 0.12$ & $2.09\pm 0.07$ \\
\hline\hline
\end{tabular}
\end{center}
\end{table}

\begin{figure}
\epsfxsize=7.0cm
\centerline{\epsffile{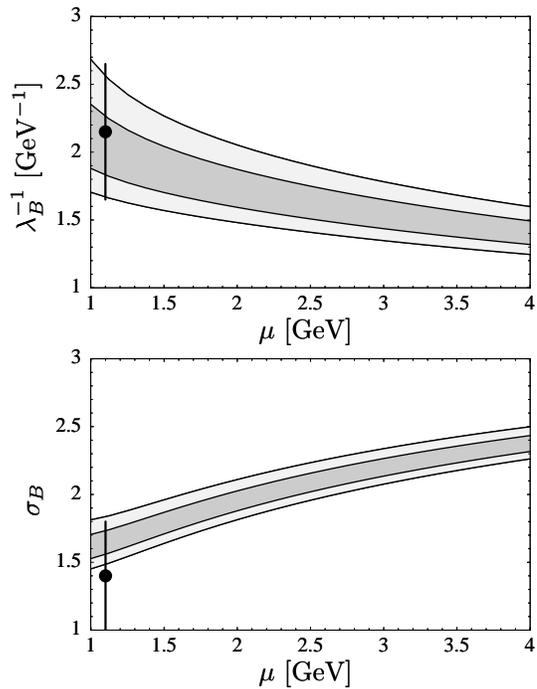}}
\caption{\label{fig:inverse} 
Model estimates of the inverse moments $\lambda_B^{-1}$ and $\sigma_B$ for 
different values of the renormalization scale. The dark bands reflect the 
uncertainty in the value of $\bar\Lambda$, whereas the light bands represent 
an estimate of the total theoretical error. The data points show the results 
obtained from the QCD sum-rule analysis of \cite{Braun:2004}.}
\end{figure}

In Table~\ref{tab:lamB} we collect the results for the two inverse moments 
obtained using the model ansatz (\ref{BLCDAmodel}). The error bars reflect 
the variation of the results with the input parameter 
$\bar\Lambda_{\rm SF}=(0.65\pm 0.06)$\,GeV. In addition, there are other 
theoretical uncertainties related to the neglect of higher-order terms in the 
OPE and, more importantly, to nonperturbative hadronic uncertainties in the 
precise shape of the LCDA for small values of $\omega$. For instance, 
comparing the results in the table with those obtained using the model 
(\ref{phiBIK}) at $\mu=1$\,GeV, we observe shifts in $\lambda_B^{-1}$ and 
$\sigma_B$ by 0.3\,GeV$^{-1}$ and 0.04, respectively. We believe that the true 
theoretical uncertainties are about twice as large as the errors shown in the 
table. A graphical representation of the results is shown in 
Figure~\ref{fig:inverse}, where the light gray bands are an estimate of the 
total theoretical uncertainty. 

Our findings are in good agreement with the QCD sum-rule estimates at 
next-to-leading order in $\alpha_s$ obtained by Braun et al.\ 
\cite{Braun:2004}, indicated by the data points in the figure. We may also 
compare with earlier estimates of $\lambda_B^{-1}$ derived from lowest-order 
QCD sum rules, where the scale dependence is not controlled. Grozin et al.\ 
\cite{Grozin:1996pq} found 
$\lambda_B^{-1}=3/(2\bar\Lambda)\approx 2.2$\,GeV$^{-1}$ (for a typical value 
$m_b\approx 4.6$\,GeV), while Ball et al.\ \cite{Ball:2003fq} obtained 
$\lambda_B^{-1}\approx 1.7$\,GeV. Both are consistent with our findings.

\section{Conclusions}

Using rigorous methods based on the operator product expansion, we have 
studied some model-independent properties of the \BM\ light-cone distribution 
amplitude $\phi_+^B(\omega,\mu)$. We have derived explicit expressions for the 
first two moments of the distribution amplitude as a function of the 
renormalization scale $\mu$ and a hard Wilsonian cutoff \UV\ applied to 
integrals over $\omega$. The ratio $M_1/M_0$ of the first two moments can be 
used to define a physical subtraction scheme for the parameter 
$\bar\Lambda=m_B-m_b$ of heavy-quark effective theory. This links the only 
nonperturbative hadronic parameter entering the moment predictions at 
next-to-leading power in $1/\UVC$ in a calculable way to the $b$-quark mass. 
From the cutoff dependence of the moment $M_0$ we have derived an analytic 
expression for the asymptotic behavior of the distribution amplitude for large
$\omega\gg\Lambda_{\rm QCD}$, valid at first order in $\alpha_s$ and at 
next-to-leading order in $1/\omega$. Finally, we have presented a new, compact
evolution formula that expresses the distribution amplitude at some scale 
$\mu$ in terms the function $\phi_+^B(\omega,\mu_0)$ at a lower scale $\mu_0$. 

Based on our analysis we have proposed a realistic model of the \BM\ 
distribution amplitude, which is consistent with the moment relations. With 
the help of this function we have obtained estimates for the inverse-moment 
parameters $\lambda_B$ and $\sigma_B$, which play an important role in many 
phenomenological applications of the QCD factorization approach to exclusive 
$B$ decays. We find $\lambda_B^{-1}=(2.1\pm 0.5)$\,GeV$^{-1}$ and 
$\sigma_B=1.6\pm 0.2$ at $\mu=1$\,GeV with conservative errors. 

We hope that our analysis will not only supply a guideline for understanding 
the \BM\ distribution amplitude without relying on a specific model, but also 
open a new strategy for further, more detailed studies of 
$\phi_+^B(\omega,\mu)$ using a systematic short-distance approach. Ultimately, 
this may help to reduce the theoretical uncertainties in predictions for 
exclusive \BM\ decays.

\vspace{0.3cm} 
{\it Acknowledgments:\/} 
We are grateful to Bj\"orn Lange and Gil Paz for useful discussions. The work 
of M.N.\ was supported in part by a Research Award of the Alexander von 
Humboldt Foundation. This research was supported by the National Science 
Foundation under Grant PHY-0355005.

\end{document}